# Nanoscale Structure and Energy Dissipation Behaviour of Tungsten Disulphide and Gold Nanoparticles and Nanoclusters Investigated by Advanced Microscopy Techniques.

**D. Devaprakasam**[*]

Department of Nanosciences and Technology, School of Nanosciences and

Technology, Karunya University, Coimbatore 641 114, India.

---

[*] e-mail:devaprakasam@karunay.edu

**Abstract**

In this study nanoscale structure and associated energy dissipation behavior of nanoparticulates of tungsten disulphide ($WS_2$) and nanoparticles of gold (Au) were investigated by advanced scanning probe and electron microscopy techniques. $WS_2$ nanoparticulates were deposited on silicon substrate by van der Waals adhesion transfer-deposition mechanisms. The templates of three size variants of Au nanoparticles, close packed (4-6nm), well dispersed (120-150nm) and clusters of (120-150 nm) nanoparticles were prepared on silicon substrates. Nano-microstructure, morphology, topography and phase shift of both $WS_2$ and Au nanoparticles and nanoparticulates were characterized by Atomic Force Microscopy (AFM), Scanning Electron Microscopy (SEM) and Transmission Electron Microscopy (TEM). SEM and TEM observations showed that the nanoparticulates of $WS_2$ are atomically oriented lamellar-hexagonal structure or inorganic graphite like structure. In the tapping mode AFM imaging, repeated scanning of nanoparticulates of $WS_2$ resulted in inter layer movement of nanolayers. It was observed that cluster of Au nanoparticles caused more phase shift and high energy dissipation compared to that of well dispersed Au nanoparticles, however close packed Au nanoparticles caused lowest phase shift and energy dissipation. Nano-microstructure, morphology, topography and phase shift were analyzed by SXM software. The nanoscale studies showed that the phase shift and the energy dissipation were influenced by the geometrical and mechanical gradient of the samples.

## 1. Introduction

Understanding and manipulating materials on the basis of requirements is a perpetual scientific endeavour to achieve the optimal usage of material, energy and time. Betterment of design and improving the efficiency of a system requires thorough knowledge of its constituents from the atomic/molecular to engineering scales. In this endeavour a lack of experimental and computational tools have till recently hampered our understanding of successfully and usefully linking properties of matter between molecular and the macro-scales. Recent development of sophisticated instruments like Atomic Force Microscope (AFM) [1] , Scanning Tunneling Microscope (STM) [2] , Scanning Electron Microscope (SEM) and Transmission Electron Microscope (TEM) [3,4] have helped to a great extent in to visualize, measure and characterize material prosperities at small length scale ~nm. New inventions and advancements in technology have given a fillip to the de-convolution of structure of materials and genesis of forces acting at the surfaces and interfaces. This has given a clear understanding of the behaviour of materials from micro scales down to atomic or molecular scales. Nanoscale interactions have profound implications on the macroscopic behaviour of materials. Different modes of behaviour at the nanoscale influence the bulk behaviour of materials. Mechanisms of nanomaterials of miniscule volumes ($nm^3$) are expected to be quite different from bulk materials because of small number of atoms or molecules in the active volume, increased surface area to volume ratio altering the thermal conductivity, and interplay between mechanical and electronic states, where continuum mechanics may breakdown. It has become increasingly important to understand and manipulate materials at the atomic and molecular scale to develop new functional materials for engineering and biomedical applications. $WS_2$ nanoparticles [5] are widely used in friction reducing agent [6] in extreme conditions and used by NASA for aerospace, military and automotive industrial applications. It is one of the most lubricous materials known to science. It can also be used in high temperature and high pressure applications. It offers temperature resistance from -270 C to 650C in normal atmosphere and from -188 C to 1316 C in vacuum. Load bearing property of coated film is extremely high at 2 GPa. It has potential application electronics device industry especially in solid –state devices.

Gold nanoparticles are important nanomaterial in arena of science and technology; it has been extensively studied due to their interesting electrical, conductive and optical properties. Gold nanoparticles and nanostructures used in high technology electronics, flexible and printable electronics [7], and electronics interconnect quantum dots and many more. Functionalized gold nanoparticles are used many biomedical applications such as targeted drug and gene delivery [8], highly sensitive diagnostic arrays [9], in-vivo photo-thermal cancer therapy treatment [10] and thermal ablation of tumors [11] by applying near-infrared radiation on gold nano-shells. In this paper nanoscale investigations of two important nanomaterials, tungsten disulphide and gold nanoparticles were studied. Nano-Microstructure, morphology, topography, phase shift, energy dissipation and associated mechanics [12-17] of the two nanomaterials have been explored.

## 2. Experimental Details

2.1. Materials

Two Ultrapure, 99.9%, Tungsten disulphide was purchased from Sky Spring Nanomaterials, USA. Gold particles suspensions of 4-6 nm and 120-150 nm with purity 99.99% were purchased from Sigma Aldrich, USA.

2.2. Specimen Preparation

For SEM and AFM analysis, nanoparticulates of $WS_2$ deposited or transferred to freshly cut and highly polished silicon substrate by van der Waals adhesion force by bringing the substrate close to nanoparticles of $WS_2$. This method was adopted to avoid any contamination from solvent. For transmission electron microscopy (TEM), the nanoparticulates of $WS_2$ collected on gold coated copper grids by the same adhesion route. The gold nanoparticles of various sizes were deposited on silicon substrate by drop coating method

2.3. Characterisation Methods

Nano-microstructure of tungsten disulphide and gold nanoparticles were characterized JOEL-JSM6500F SEM, which has resolution of about 3 nm. Nanostructures of tungsten disulphide was characterised by JOEL 2010 TEM at 200 kV. It is capable of point-to-point resolution of 0.19 nm, with the ability to image lattice fringes at 0.14 nm resolution or better. Tapping Mode AFM measurements were carried out using a Nanoscope IIIa AFM and a MPP-11100 rectangular phosphorous doped silicon cantilever of stiffness 40 N/m with nominal tip radius ~ 10 nm and resonant frequency of ~356 kHz.

## 3. Results

3.1. Nano/Microstructure characterization of tungsten disulfide ($WS_2$)

$WS_2$ is layered hexagonal structure with easy shear direction, c-axis d-spacing (0 0 2) is 6.2Å, a sheet of tungsten atoms is sandwiched between two hexagonally packed sulphur layers [5, 6, 18]. Schematic atomic arrangement of close packed layered hexagonal and sandwich structure of $WS_2$ is shown in Fig. 1. The bonding within the S–W–S sandwich is covalent, while weak van der Waals forces hold the sandwich together resulting in inter-lamellar mechanical weakness. Thus, under a shearing force the basal planes slide back and forth over one another by intra-crystalline slip.

Nano-micro structure investigations by high resolution SEM revealed nested clusters structure (Fig.2a), layered sturcture (Fig.2b), steps in the layered structure (Fig.2c) and rolled layer structure (Fig. 2d) of $WS_2$. Fig. 2a-d shows that the platelets or nanoparticulates of $WS_2$ varies from 500 nm to 1 μm and also there are large platelets of $WS_2$ of few μm sizes. In the case less than 1 μm size, the edges are curved and spline shape, however when the particles are above 1 μm the edges of platelets are cleaved in polygonal shapes with clear steps and layer arrangment. It is important to note that few atoms thick layer can be rolled and could form $WS_2$ nanotubes. Partially rolled layer of $WS_2$ is shown in Fig. 2d.

High resolution TEM image of $WS_2$ (Fig. 3a-b) shows the hexagonal lattice arrangement of the tungsten and sulfur atoms. FFT image of ($WS_2$) (Fig. 3b) confirm the hexagonal lattice arrangement.

3.2. Energy Dissipation of tungsten disulfide ($WS_2$)

Nanostructure and nanoparticles behavior during external force field is very important for the nanomanipulation. Any dynamical changes in the morphology affects the phase shift and energy disspation. A simple nanomanipulation studies of 60 nm $WS_2$ nanoparticulates was carried out using AFM in the tapping mode operation.

During the first scan (Fig. 4a-b), the layered nanoparticle is very stable. The line profile across the $WS_2$ nanoparticle shows that the height is about 20 nm and width is about 60 nm. It was observed that at 20 nm of width from the edge there is a small step of height 10 nm. The phase image and its corresponding line shows the phase shift at various position along the $WS_2$ nanoparticles.

The changes in the phase shift associated with energy dissipation due to tip-sample interaction. When the tip scan the sample, at the silicon- $WS_2$ boundary (Fig. 4 a-b) there is lot energy dissipated due to inelastic collision of tip on the sample, which results in large fluctuation of phase shift (Fig. 4c-d). I also observed similar phase shift fluctuation at the inter layer boundary (Fig. 4b and 4d). The inelastic collision of the tip with sample, phase shift and therefore energy dissipation are very much influenced by geometrical gradient and mechanical modulus gradient [12-17]

After few repeated scan, the top layer is moved relative to the bottom layer which is shown in Fig. 5a-d. The line profile of the topography image clearly shows the edge of the layer top layer at 15 nm and the corresponding height is increased to 30 nm due to inter layer gap and rearrangement. The fluctuation in the phase shift becomes more prominent at the interlayer boundary and edge of the second layer. The line profile of the phase shift shows high fluctuation at the edge of top layer (Fig.5d).

Now it is clear that both geometrical gradient and mechanical modulus gradient have greater influence in the inelastic collision and phase shift, any dynamical changes due to interlayer movement affects over all energy dissipation. Furthermore, $WS_2$ nanoparticles can be manipulated with small oscillation force and layers can be rearranged. With direction of oscillation force, we can manipulate the inter layer arrangement as it is required for the application.

3.3. Nano/Microstructure and Energy dissipation of gold nanoparticles and clusters (Au)

Gold Nanopartcles (Au): For centuries gold has been used by us for various purposes including ornaments making and medicinal applications [7-11]. Gold nanoparticles have been extensively studied for their unique optical, electronic and molecular recognition properties with applications in electronics, targeted drug delivery, photo-thermal ablation of tumours. Gold nanoparticles are prepared by various chemical and physical routes [19, 20]. It can be produced in different shapes and sizes which in fact influence properties and applications

Fig. 6a and 6c show the AFM tapping mode image of 4-6 nm close packed gold nanoparticles, in which a region was identified which contains second layer of gold nanoparticles on the first layer of close packing. The line profile shows that the second layer has height about 4 nm and width is about 30 nm.

Fig. 6 c –d show that the first layer of close packing dissipates very negligible amount energy which very evident from the associated phase shift (Fig. 6d). This is due to homogeneity of close packing and modulus properties i.e. negligible geometrical and modulus gradient.

However phase shift fluctuation significantly more in the second close packed layer due to 4 nm step geometrical gradient. Overall energy dissipation of both first and second layer close packed gold nanoparticles assembly is negligible.

Similar studies were carried out on well dispersed 120-150 nm gold nanoparticles on silicon substrate (Fig. 7a-d). The line profile of the two nanoparticle (Fig. 7b) shows that the width is about 120-150 nm and the height is about 16 nm. The discrepancy in height value may be due to faceted nature of gold nanoparticle or any repulsive force acted between tip and gold nanoparticle or the gold nanoparticle may be sitting on some nano trenches of silicon substrate. Fig.7d shows the line profile of the phase image, high phase shift and energy dissipation at the boundary of gold nanoparticle and silicon substrate. Therefore geometrical and modulus gradient causes more phase shift and energy dissipation.

Further studies of clusters of the gold nanoparticles on the silicon substrates were done. Fig. 8a-d show topography and phase image of the gold nanoparticles clusters and their line profiles. The maximum height of the clusters is about 250 nm and width is about 1µm. the line profile of the phase image shows very high phase shift fluctuation due to in-elastic collision tip on the nanoparticles. Variation of force fields around nanoparticles step also introduces fluctuation, geometrical and modulus gradient play very significant role in the overall phase shift changes and energy dissipation mechanisms

## 4. Discussion

In this studies I have explored and investigated two different nanoparticles, $WS_2$ and gold, which have different morphology, nano-microstructure and properties. $WS_2$ particles are layered hexagonal close-packed structure, they have easy shear plane in the c-axis between S-W-S sandwiches. This is due to very weak van der Waals interactions between them. For that reason $WS_2$ is very lubricious materials. It can be used in coating MEMS devices, storage and memory devices in order to achieve smooth relative motion of dynamic structure. The nanoparticles of $WS_2$ can be rearranged to achieve different configuration by oscillating or moving tip. By systematic nanomanipulation, even mono atomic layer of $WS_2$ is achievable and it can be rolled to get hollow nanotubes for electronics and sensor applications. I also found that geometrical and mechanical modulus gradient causes phase shift and energy dissipation. Further investigations are required to achieve different configuration and arrangement of $WS_2$ layers, I am exploring various possibilities and it will be reported in the future communications. I also carried out nanoscale investigations of gold nanoparticles in different arrangement and configurations to understand phase shift and energy dissipation behaviour due to geometrical and mechanical modulus gradient. Close packed 4-6 nm gold nanoparticles templates show less phase shift and energy dissipation. However well dispersed individual nanoparticles show high phase shift and energy dissipation, it is more in the case of clustered nanoparticles. I explored the reasons of this characteristic using AFM dynamics and deformation mechanics Both geometrical and mechanical modulus gradient causes fluctuations in force fields between tip and sample, any dynamical changes in geometry or mechanical modulus introduce fluctuations in the force field which in turn introduces chaotic motion of the tips and local material deformations, dissipative inelastic collisions result in high phase shift and energy dissipation. Introducing gradual and infinitesimal variations in property gradient could reduce the dissipative process and to achieve smooth dynamical motion and less energy dissipation.

## 5. Acknowledgments

I thank and appreciate School of Nanosciences and Technology, Karunya University, Coimbatore; Mechanical Engineering, Indian Institute of Science, Bangalore; Engineering Materials, University of Sheffield, UK and UKIERI programme, DST-British Council for their support. I dedicate this research work to Prof. S.K.Biswas (1945-2013), Mechanical

Engineering, Indian Institute of Science, Bangalore, who inspired me and guided me as a research supervisor.

## 6. References

[1]    Binning, G., Quate, C.F., and Gerber, Ch.: Atomic Force Microscope, Phys. Rev. Lett. 56 930-933, (1986).

[2]    Binning, G., Quate, C.F., and Weibel, E.:Surface Studies by Scanning Tunnelling Microscopy, Phys. Rev. Lett. 49, 57-61 (1982).

[3]    Williams, D.B., and Carter, C. B.: Transmission Electron Microscopy, 4$^{th}$ ed, Plenum Press, New York, (1996).

[4]    Reimer, L.: Transmission Electron Microscopy, 2$^{nd}$ ed, Springer, (1997).

[5]    Scharf, T.W., Prasad, S.V., Dugger, M.T., Kotula, P.G., Goeke, R.S., Grubbs, R.K.: Growth, structure, and tribological behavior of atomic layer-deposited tungsten disulphide solid lubricant coatings with applications to MEMS, Acta Materialia 54, 4731–4743 (2006).

[6]    Scharf, T.W., Prasad, S.V., Mayer, T.M., Goeke, R.S., and Dugger, M.T.: Atomic layer deposition of tungsten disulphide solid lubricant thin films, J. Mater. Res. 19, 3443-3446 (2004).

[7]    Hu, L., Kim, H. S., Lee, J.-Y., Peumans, P., Cui, Y.: Scalable Coating and Properties of Transparent, Flexible, Silver Nanowire Electrodes. ACS Nano. 4 2955–2963 (2010).

[8]    Han, G., Ghosh, P., and Rotello, V. M.: Functionalized gold nanoparticles for drug delivery. Nanomedicine. 2, 113-123 (2007).

[9]    Fourkal, E.; Velchev, I.; Taffo, A.; Ma, C.; Khazak, V.; Skobeleva, N.: Photo-Thermal Cancer Therapy Using Gold Nanorods," World Congress on Medical Physics and Biomedical Engineering 2009, IFMBE Proceedings. 25, 761-763 (2009).

[10]   Lukianova-Hleb, E. Y.; Hanna, E. Y., Hafner, J. H., Lapotko, D. O.: Nanotechnology. 21, 085102 (2010).

[11]   Weisenhorn, A.L., Hansma, P.K., Albrecht, T.R., and Quate, C.F.: Forces in atomic force microscopy in air and water, Appl. Phys. Lett. 54, 2651-2653 (1989).

[12]   Capella, B., Baschieri, P., Frediani, C., Miccoli, P. and Ascoli,C.: Force-Distance Curves by AFM: A Powerful Technique for Studying Interactions. IEEE Engineering in medicine and biology, 58-65 (1997).

[13]   Weisenhorn, A.L., Maivald, P., Butt, H.J., and Hansma, P.K.: Measuring adhesion, attraction, and repulsion between surfaces in liquids with an atomic-force microscope, Phys. Rev B. 45, 11226-11232 (1992).

[14]   Cappella, B. and Dietler, G.: Force-distance curves by atomic force microscopy, Surface Science Reports 34 1-104 (1999).

[15]   Ruan, J.A., and Bushan, B.: Atomic-Scale Friction Measurement Using Friction Force Microscopy: Part I-General Principles and New Measurement Techniques Trans. ASME, J.Tribol. 116, 378-389 (1996).

[16]   Garcia, R.: Indentification of nanoscale dissipation process by dynamic atomic force microscopy. Phys Rev Lett 97, 016103 (2006) .

[17]   Garcia, R., Magerle, R., Perez, R.: Nanoscale compostional mapping with gentle forces. Nat. Mater. 6, 405-411 (2007).

[18]   Prasad, S., and Zabinski, J.: Lubricants: Super slippery Solids. Nature. 387, 761-763 (1997).

[19]   Fumitaka Mafuné, Jun-ya Kohno, Yoshihiro Takeda, and Tamotsu Kondow, Hisahiro Sawabe.: Formation of Gold Nanoparticles by Laser Ablation in Aqueous Solution of Surfactant. J. Phys. Chem. B. 105, 5114–5120 (2001).

[20]   Martin, M.N., Basham, J.I., Chando, P., Eah, S.-K.: Charged gold nanoparticles in non-polar solvents: 10-min synthesis and 2D self-assembly. Langmuir. 26, 7410 (2010) .

## 7. Figures

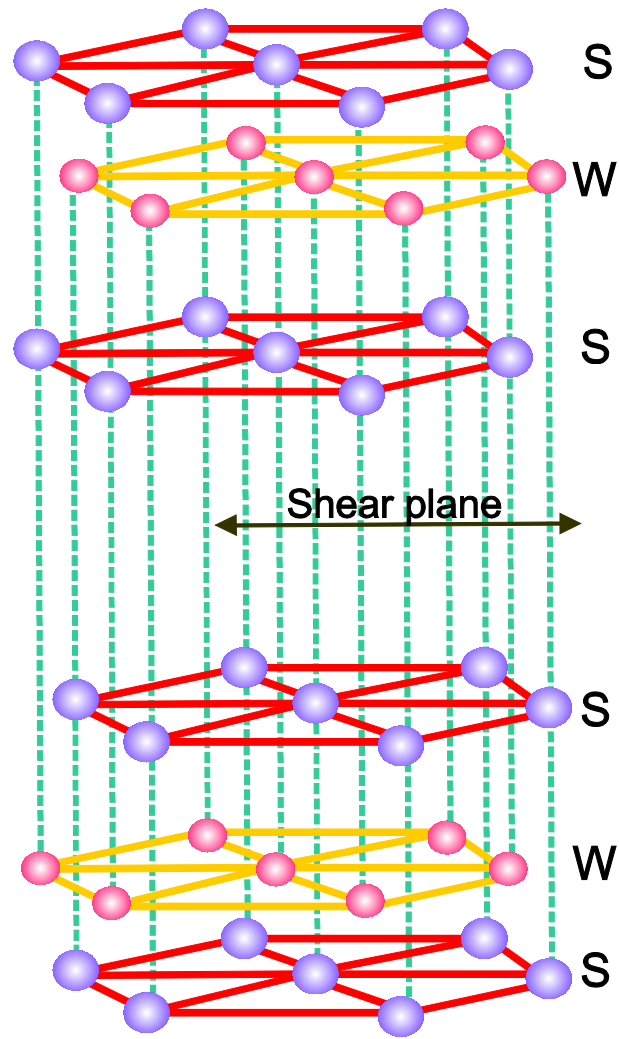

Figure 1. Structure of Tunsten disulphide, S represents sulphur atoms, W represents tungsten, tunsten hexogonal layer is sandwiched between two hexagonal layer of sulphur. The c-axis d-spacing between two layers S-W-S sandwich is 6.2Å.

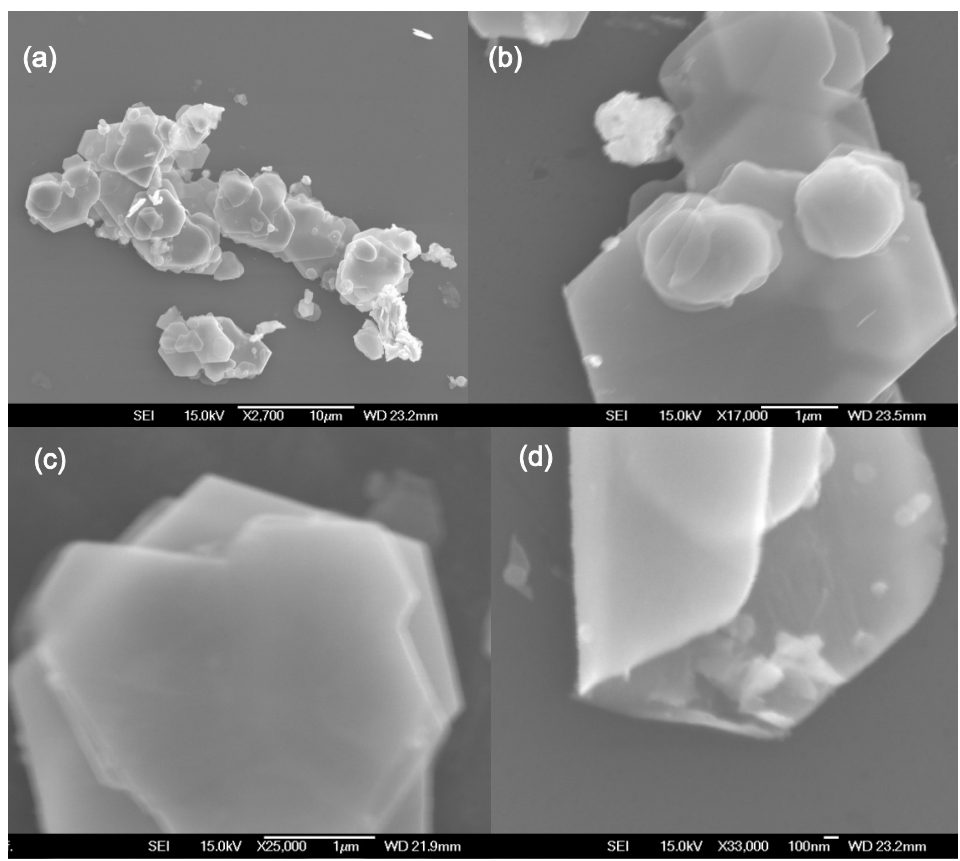

Figure 2. SEM images of tungsten disulphide (a) nested clusters of $WS_2$ (b) different morphology of ($WS_2$) (c) layered structure of $WS_2$ (e) partially rolled layer of $WS_2$

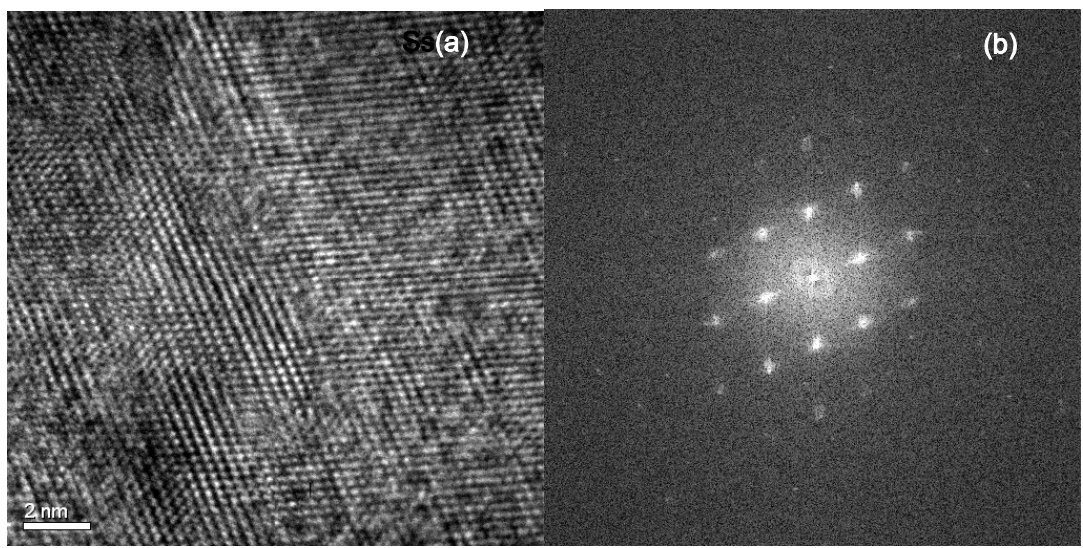

Figure 3. TEM image of tungsten disulphide (a) hexagonal close packed structure of $WS_2$ (b) FFT image of hexagonal close packed structure.

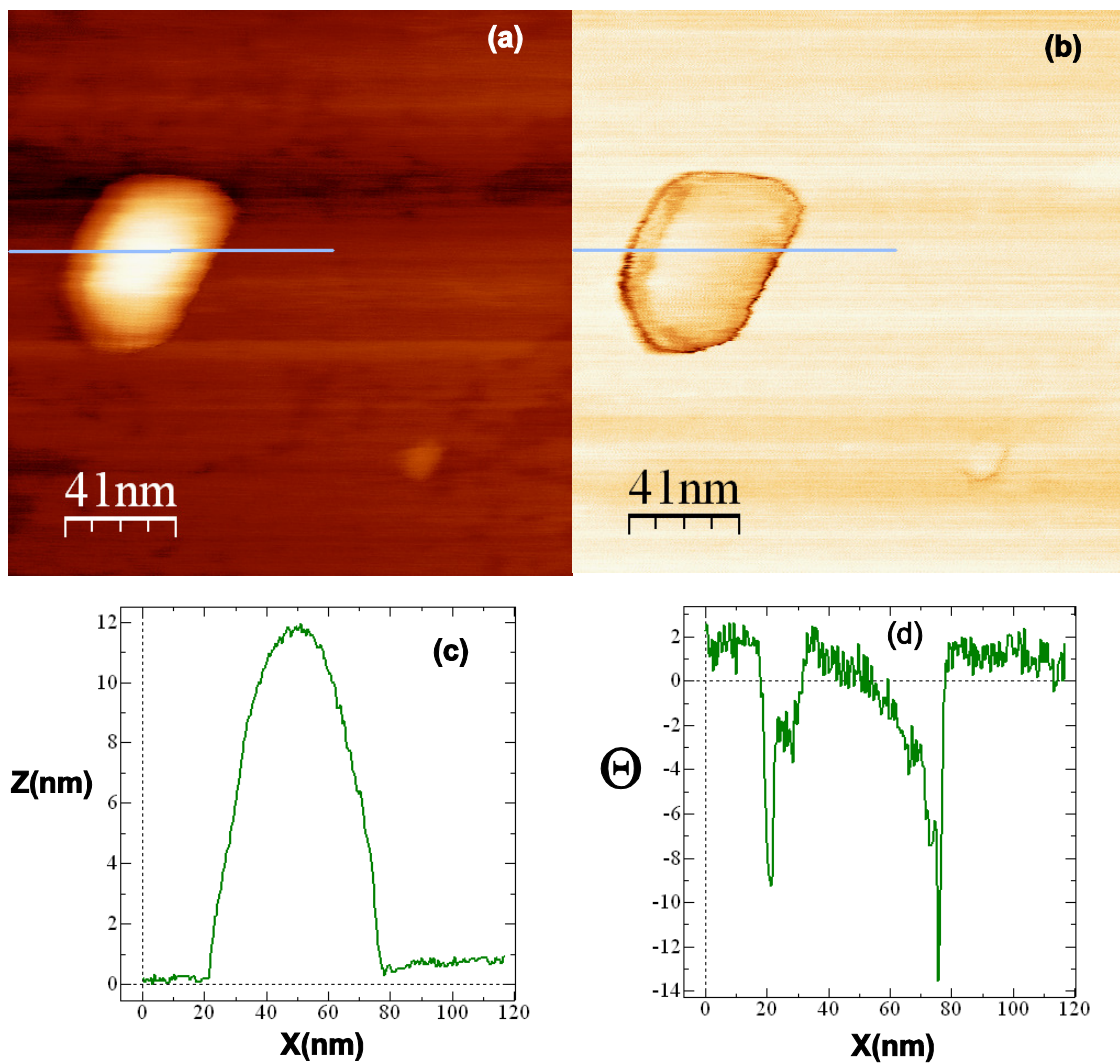

Figure. 4. AFM tapping mode images of a nanoparticles of tungsten disulphide nanoparticles (a) topography image (b) phase image (c) line profile of the topography image (d) line profile of the phase image.

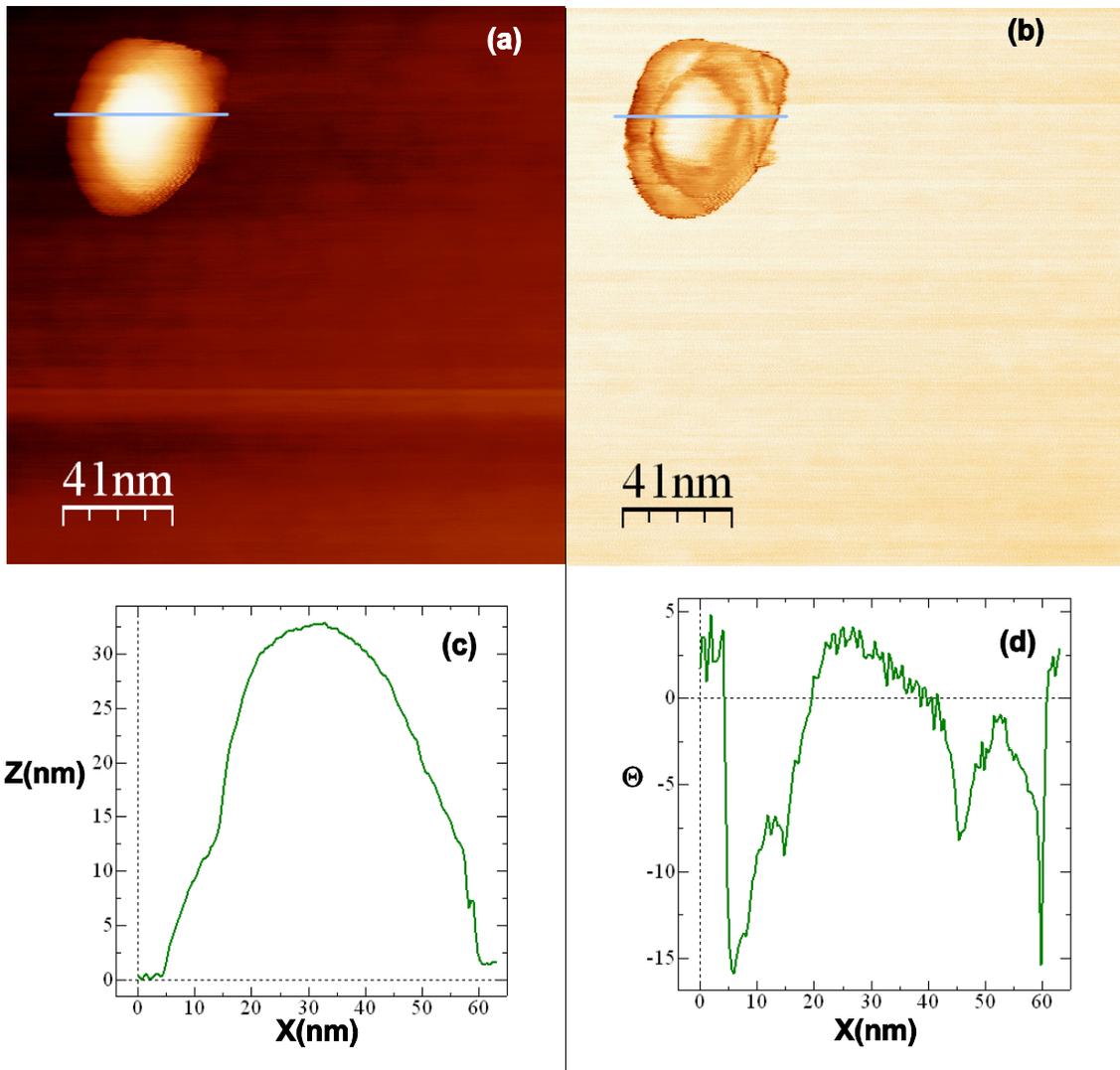

Figure. 5. AFM tapping mode images of a tungsten disulphide nanoparticle were taken after few scans (a) topography image (b) phase image (c) line profile of the topography image (d) line profile of the phase image.

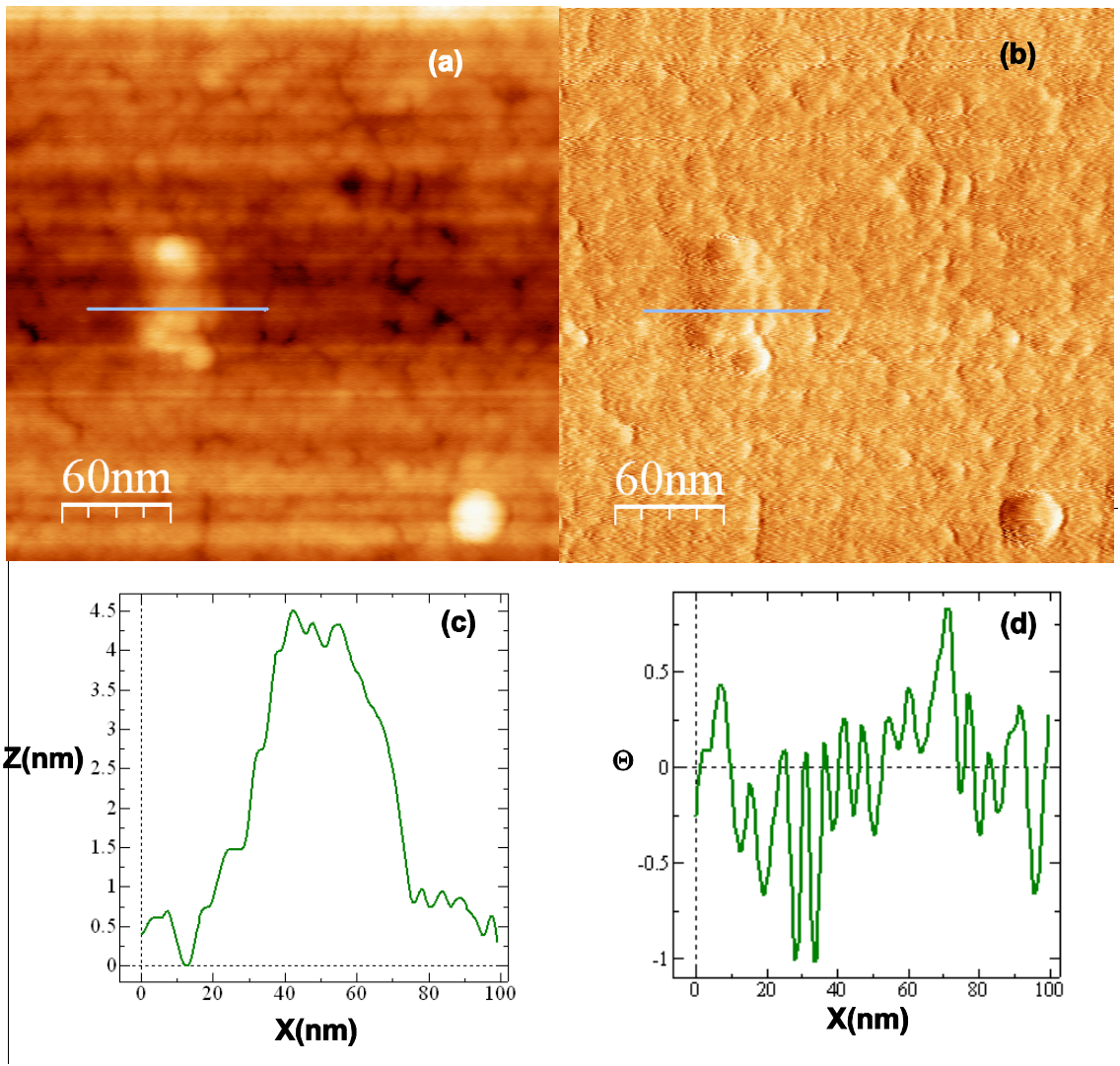

Figure. 6. AFM tapping mode images of the closed Au nanoparticles template (a) topography image (c) phase image (c) line profile of the topography image (d) line profile of the phase image.

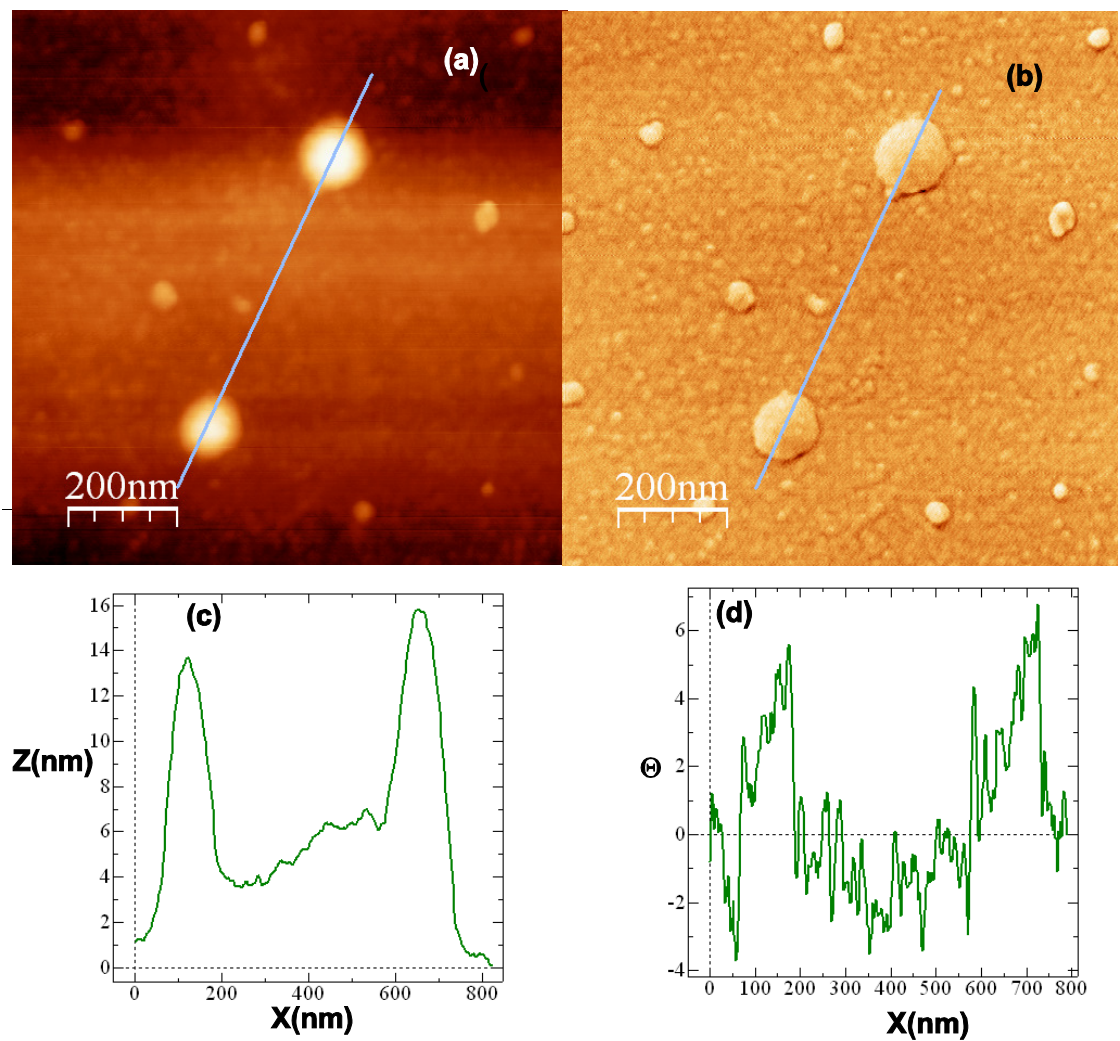

Figure. 7. AFM tapping mode images of the Au nanoparticles (a) topography image (b) phase image (c) line profile of the topography image (d) line profile of the phase image.

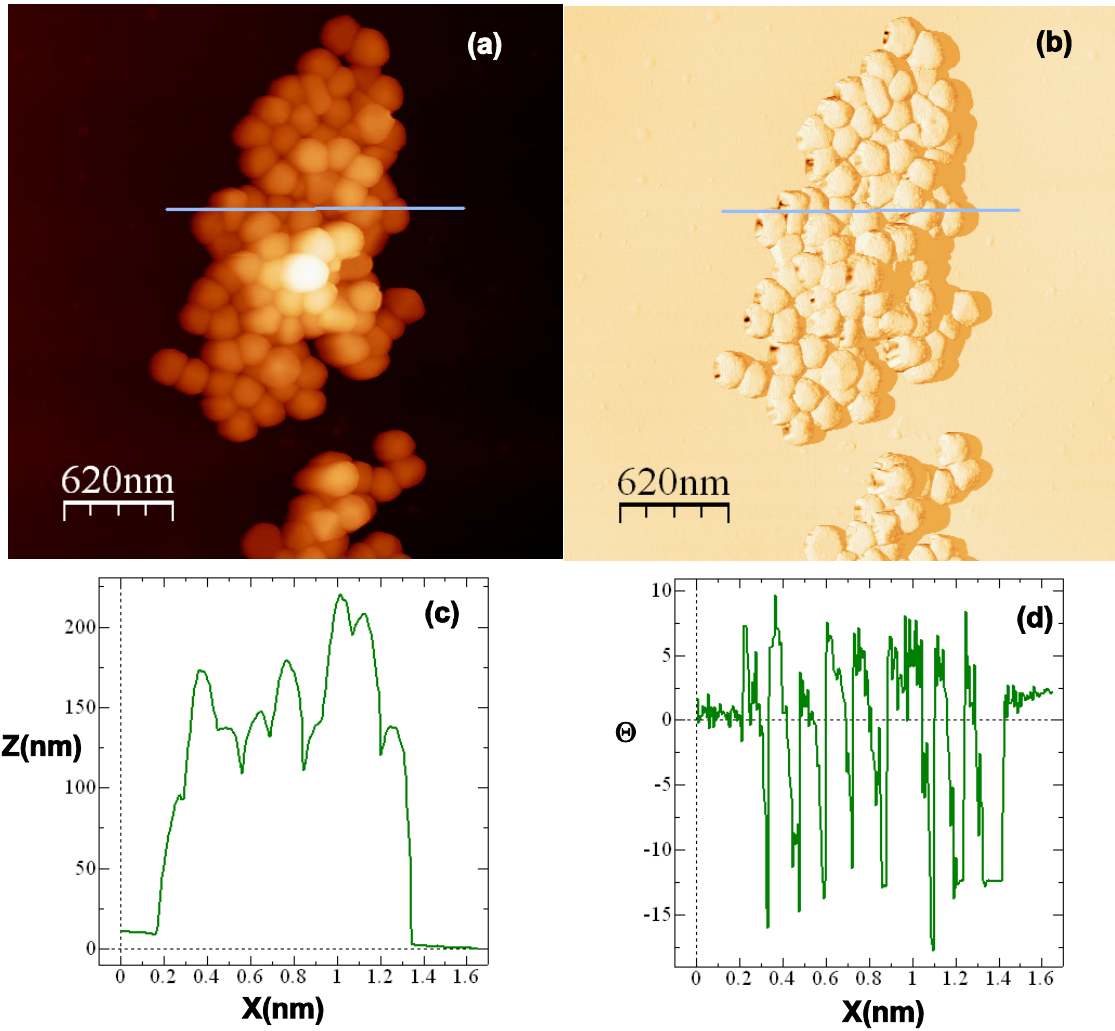

Figure. 8. AFM tapping mode images of the Au nanoclusters (a) topography image (b) phase image (c) line profile of the topography image (d) line profile of the phase image.